\documentclass[11pt]{article}
\usepackage[utf8]{inputenc}
\usepackage[T1]{fontenc}
\usepackage{amsmath,amssymb,amsfonts,amsthm}
\usepackage{physics} 
\usepackage{graphicx}
\usepackage{geometry}
\usepackage[colorlinks=true, pdfstartview=FitV, linkcolor=blue,citecolor=blue, urlcolor=blue]{hyperref}
\usepackage{xcolor}
\usepackage{authblk}
\usepackage[numbers,sort&compress]{natbib} 
\usepackage{xcolor}

\title{Solving nonlinear subsonic compressible flow in infinite domain \\
via multi-stage neural networks}
\date{} %

\author[1, 2,$^{\dagger}$]{Xuehui Qian}
\author[3, $^{\dagger}$]{Hongkai Tao}
\author[4, 5, *]{Yongji Wang}

\affil[1]{\small Washington University in St. Louis, Department of Mechanical Engineering and Material Science, St. Louis, MO 63130, USA}
\affil[2]{\small University of Notre Dame, Department of Aerospace and Mechanical Engineering, Notre Dame, IN 46556, USA}
\affil[3]{\small Central South University, Dundee International Institute,  Hunan, 410083, China}
\affil[4]{Stanford University, Department of Geophysics, Stanford, CA 94305, USA}
\affil[5]{New York University, Department of Mathematics, New York, NY 08550, USA}
\affil[$^{\dagger}$]{These authors contribute equally}
\affil[*]{Corresponding authors: yongjiw@stanford.edu}

\begin{document}

\maketitle

\begin{abstract}
In aerodynamics, accurately modeling subsonic compressible flow over airfoils is critical for aircraft design. However, solving the governing nonlinear perturbation velocity potential equation presents computational challenges. Traditional approaches often rely on linearized equations or finite, truncated domains, which introduce non-negligible errors and limit applicability in real-world scenarios. In this study, we propose a novel framework utilizing Physics-Informed Neural Networks (PINNs) to solve the full nonlinear compressible potential equation in an unbounded (infinite) domain. We address the unbounded-domain and convergence challenges inherent in standard PINNs by incorporating a coordinate transformation and embedding physical asymptotic constraints directly into the network architecture. Furthermore, we employ a Multi-Stage PINN (MS-PINN) approach to iteratively minimize residuals, achieving solution accuracy approaching machine precision. We validate this framework by simulating flow over circular and elliptical geometries, comparing our results against traditional finite-domain and linearized solutions. Our findings quantify the noticeable discrepancies introduced by domain truncation and linearization, particularly at higher Mach numbers, and demonstrate that this new framework is a robust, high-fidelity tool for computational fluid dynamics.\\
\\
{\bf keywords}: subsonic flow; infinite domain; physics-informed neural networks; multi-stage training
\end{abstract}

\maketitle

\renewcommand\theequation{\arabic{section}.\arabic{equation}}

\section{Introduction}

Subsonic compressible flow is crucial in the context of aerodynamics due to its wide application in aircraft design, where understanding pressure distribution and aerodynamic loading is essential \cite{anderson2011ebook,anderson1990compressible,liepmann1957,shapiro1953}. To understand its behavior over airfoils, a governing equation, called the perturbation velocity potential equation, has been derived \cite{ashley1965aero, anderson2011ebook}. Traditionally, this equation has offered valuable insights into fluid dynamics, facilitating the prediction of aerodynamic forces and the optimization of airfoil shapes for enhanced efficiency and lift characteristics \cite{anderson2011ebook, abbott1959}. However, historically, the velocity potential equations presented a significant challenge because they are nonlinear partial differential equations, and hence there was no general closed-form solution except under restrictive assumptions \cite{ferri1976, hayesprobstein1959}. A classical method to address this challenge is to seek assumptions based on the flow physics and linearize the equations \cite{anderson1990compressible}. Despite the fact that analytic or semi-analytic solutions could be obtained by conventional mathematical techniques, they lose accuracy when nonlinear compressibility effects become significant \cite{lighthill1951, vandyke1975}.

The linear solutions dominated the history of aerodynamics up to the mid-1950s. Since then, with the appearance of the high-speed computer, computational fluid dynamics (CFD) has thrived, providing numerical methods to solve nonlinear equations more accurately \cite{anderson2002computational, hirsch1988, pletcher2012computational}. Currently, with finite-difference, finite-volume, and finite-element methods, we can solve the nonlinear equations numerically with high fidelity \cite{johnson1987fem, randall2007fdm, moukalled2015fvm, leveque2002}. However, the core of most CFD programs is a linear algebra solver, which means that for nonlinear equations, it is necessary to discretize and handle part of the nonlinear problem within linear algebra procedures \cite{blazek2015}. More importantly, in practice, unbounded domains are truncated to finite computational domains with approximate outer boundary conditions. Therefore, despite the improvement in accuracy compared to the linear theory, truncation and artificial outer boundaries can still introduce noticeable errors when high-fidelity solutions in effectively infinite domains are required.

With the development of machine learning, an innovative method called Physics-Informed Neural Networks (PINNs) is rapidly evolving \cite{lagaris1998artificial, raissi2019physics, yu2022gradient, WANG2024112865}. It represents a paradigm shift in tackling such nonlinear partial differential equations and has proven its advantages in solving these problems across various fields \cite{cai2021physics, haghighat2021physics, jagtap2022physics, cao2024solver}. PINNs are essentially deep neural networks (DNNs). Conventional DNNs do not incorporate physical laws and require a large amount of data for the training. However, by taking the equations that govern the flow as loss functions, PINNs successfully take the physical laws into consideration. In addition, PINNs enable the acquisition of solutions from the original trivial solution by solely enforcing physical laws via soft constraints in the loss function instead of depending on massive amounts of data, which is more efficient compared with standard DNNs \cite{karniadakis2021physics}. Compared with traditional numerical methods, PINNs exhibit several advantages. Primarily, PINNs do not require meshing and thus can handle high-dimensional problems and complex geometries with small-scale features \cite{abueidda2021meshless}. They also avoid explicit discretization, as partial derivatives can be computed directly. Additionally, PINNs exhibit the ability to solve infinite domain problems, which are regarded as a challenge for standard numerical methods \cite{moseley2023finite, xia2023spectrally, ren2024seismicnet, wang2025high, bararnia2022application}. However, standard PINNs often have limited accuracy for challenging nonlinear PDEs, and solving problems in an infinite domain remains difficult, requiring careful treatment. Motivated by these advantages, we adopt PINNs as the base solver and develop advanced settings to improve accuracy for nonlinear problems and handle the infinite domain in subsonic flow.

While our motivation comes from airfoil aerodynamics, we develop and validate our method using simplified geometries, namely a circular cylinder and an ellipse. The remainder of this paper is organized as follows. Section~\ref{sec:Mathematical} establishes the mathematical framework, while Section~\ref{sec:methods} outlines the foundational methodology for PINNs. Section~\ref{sec:advance} begins by quantifying the significant errors introduced by finite-domain truncation. To address this, we introduce a methodology that incorporates an infinite-domain coordinate transformation, physical asymptotic constraints, and the Multi-Stage PINN (MS-PINN) approach, followed by validation against the incompressible flow benchmark. Section~\ref{sec:results} extends the framework to compressible flow, investigating both linear and nonlinear equations. This section analyzes the discrepancies inherent in linearization methods and demonstrates the framework's versatility by solving for flow over an ellipse. Finally, Section~\ref{sec:conclusion} provides concluding remarks.

\section{Mathematical setting of compressible subsonic flow}\label{sec:Mathematical}

\subsection{Governing Equations}\label{sec:goveq}
In this study, we focus on a two-dimensional, steady, compressible subsonic flow over an elliptical geometry immersed in a uniform stream. To describe the flow, we begin with some assumptions: the flow is inviscid and irrotational. This allows us to introduce the velocity potential $\Phi=\Phi(x,y)$ for the flow as $\textbf{V} = \nabla\Phi$, namely:
\begin{gather}\label{eq:vpotential}
    u = \frac{\partial\Phi}{\partial x} \quad \mathrm{and} \quad
    v = \frac{\partial\Phi}{\partial y},
\end{gather}
where $u$ and $v$ are the velocity components in $x$ and $y$ directions, respectively.

The motion of compressible flow is governed by the Navier-Stokes equations and the continuity equation, which describe the conservation of momentum and mass, respectively. For two-dimensional compressible flow, the continuity equation is given by:
\begin{gather}\label{eqn:conti}
    \frac{\partial\rho}{\partial t}+\nabla\cdot\left(\rho\textbf{u}\right)=0,
\end{gather}
where $\rho$ represents the air density and ${\bf u} = (u, v)$ is the velocity vector. Since we are assuming steady flow, the time derivative term disappears. Next, by substituting equation~\eqref{eq:vpotential} into the continuity equation~\eqref{eqn:conti}, we obtain:
\begin{gather}\label{eq:origin}
    \rho\left(\frac{\partial^2\Phi}{\partial x^2}+\frac{\partial^2\Phi}{\partial y^2}\right)+\frac{\partial\Phi}{\partial x}\frac{\partial\rho}{\partial x}+\frac{\partial\Phi}{\partial y}\frac{\partial\rho}{\partial y}=0.
\end{gather}

Similarly, the Navier-Stokes equations describe the balance of momentum and are written in vector form as:
\begin{gather}\label{eq:NS}
    \rho\frac{\partial\mathbf{u}}{\partial t} + \rho\left(\mathbf{u}\cdot\nabla\right)\mathbf{u} = -\nabla\mathbf{p} + \mathbf{F}_{\nu} + \rho\mathbf{f},
\end{gather}
where $\mathbf{p}$, $\mathbf{F}_{\nu}$, and $\mathbf{f}$ represent pressure, viscous stress, and the body forces per unit mass, respectively.  Since the flow is steady, the time derivative term is omitted. Furthermore, under the assumption of inviscid flow, the viscous stress term becomes negligible, i.e., $\mathbf{F}_{\nu} \approx 0$. Additionally, gravity effects (body forces) for the airflow are negligible in this study, so the momentum equation simplifies to:
\begin{gather}
    \rho\left(\mathbf{u}\cdot\nabla\right)\mathbf{u} = -\nabla\mathbf{p} \qquad \Longrightarrow \qquad \frac{\rho}{2} \, \nabla \left({\bf u}^2 \right) =-\nabla\mathbf{p},
\end{gather}
where $\mathbf{u}^2 = u^2 + v^2$. 
Substituting equation~\eqref{eq:vpotential} into the above equation, we obtain:
\begin{gather}\label{eq:eqnp}
    dp=-\frac{\rho}{2}d\left[\left(\frac{\partial\Phi}{\partial x}\right)^2+\left(\frac{\partial\Phi}{\partial y}\right)^2\right].
\end{gather}
Since the flow is isentropic, we can use isentropic relations, which directly link changes in pressure, $dp$, to changes in density, $d\rho$. Hence, we have the relation that $\frac{dp}{d\rho}=\left(\frac{\partial p}{\partial\rho}\right)_s$. Also, the right-hand side is the square of the speed of sound, denoted by $a^2$. Therefore, we have:
\begin{gather}\label{eq:eqna}
    dp=a^2d\rho.
\end{gather}
Substituting equation~\eqref{eq:eqna} into equation~\eqref{eq:eqnp}, we obtain:
\begin{gather}\label{eq:eqn0}
    d\rho=-\frac{\rho}{2a^2}d\left[\left(\frac{\partial\Phi}{\partial x}\right)^2+\left(\frac{\partial\Phi}{\partial y}\right)^2\right].
\end{gather}

Combining equations \eqref{eq:eqn0} and \eqref{eq:origin}, we obtain the final governing equation for the velocity potential of compressible subsonic flow as:
\begin{gather}\label{eq:eqn1}
    \left[1 - \frac{1}{a^2}\left(\frac{\partial\Phi}{\partial x}\right)^2\right]\frac{\partial^2\Phi}{\partial x^2} + \left[1 - \frac{1}{a^2}\left(\frac{\partial\Phi}{\partial y}\right)^2\right]\frac{\partial^2\Phi}{\partial y^2} - \frac{2}{a^2}\left(\frac{\partial\Phi}{\partial x}\right)\left(\frac{\partial\Phi}{\partial y}\right)\frac{\partial^2\Phi}{\partial x\partial y} = 0.
\end{gather}
Next, we introduce the perturbation velocity potential $\phi$. Specifically, for the body immersed in a uniform flow with velocity $U_\infty$ along the $x$ direction, we decompose the velocity potential as
\begin{gather}\label{eq:decompose}
    \Phi(x,y) = U_\infty x + \phi(x,y).
\end{gather}
Substituting this relationship into equation~\eqref{eq:eqn1}, we arrive at the nonlinear perturbation velocity potential equation:
\begin{gather}\label{eq:eqn1_pert}
    \left[1 - \frac{1}{a^2}\left(\frac{\partial\phi}{\partial x}+U_\infty\right)^2\right]\frac{\partial^2\phi}{\partial x^2} + \left[1 - \frac{1}{a^2}\left(\frac{\partial\phi}{\partial y}\right)^2\right]\frac{\partial^2\phi}{\partial y^2} - \frac{2}{a^2}\left(\frac{\partial\phi}{\partial x}+U_\infty\right)\left(\frac{\partial\phi}{\partial y}\right)\frac{\partial^2\phi}{\partial x\partial y} = 0.
\end{gather}

For steady, adiabatic, inviscid flow, the energy equation is written as:
\begin{gather}\label{eq:energy}
    h_1 + \frac{V_1^2}{2} = h_2 + \frac{V_2^2}{2},
\end{gather}
where $h$ is the enthalpy, and for a calorically perfect gas, $h = c_{p}T$ with $c_p = \frac{\gamma R}{\gamma-1}$ being the specific heat at constant pressure. Substituting these relations into equation~\eqref{eq:energy}, we obtain:
\begin{gather}\label{eq:energy2}
    \frac{\gamma RT_1}{\gamma-1} + \frac{V_1^2}{2} = \frac{\gamma RT_2}{\gamma-1} + \frac{V_2^2}{2}.
\end{gather}
Since the speed of sound is given by $a = \sqrt{\gamma RT}$, equation~\eqref{eq:energy2} can be rewritten as:
\begin{gather}
    \frac{a_1^2}{\gamma-1}+\frac{V_1^2}{2} = \frac{a_2^2}{\gamma-1}+\frac{V_2^2}{2}.
\end{gather}
Taking $a_1$ as $a$ and $a_2$ as the stagnation sound speed, we obtain:
\begin{gather}\label{eq:eqn2}
    a^2 = a^2_0 - \frac{\gamma - 1}{2}\left[\left(\frac{\partial\phi}{\partial x}+U_\infty\right)^2 + \left(\frac{\partial\phi}{\partial y}\right)^2\right],
\end{gather}
where $a_0$ is a known constant and $\gamma$ is the heat capacity ratio. In this study, we consistently use $\gamma = 1.4$.

\subsection{Linearization}\label{sec:linearization}
The nonlinear perturbation velocity potential equation presented earlier is challenging to solve analytically. In classical aerodynamics, small-disturbance assumptions are commonly used to linearize the governing equations, making them easier to solve. However, in this study, we do not use linearization as a method to simplify or approximate the nonlinear problem. Instead, the linearized equation motivates the design of our method and serves as a reference for evaluating the performance of our model.

Based on~\eqref{eq:decompose} and the definitions $u=\Phi_x$ and $v=\Phi_y$, we define the perturbation velocities as
\begin{gather}\label{eq:uvphi}
    u'(x,y) = \phi_x(x,y), \qquad v'(x,y) = \phi_y(x,y),
\end{gather}
so that $u = U_\infty + u'$ and $v = v'$.

By substituting \eqref{eq:uvphi} into \eqref{eq:eqn1_pert}, and using $\phi_{xx}=u'_x$, $\phi_{yy}=v'_y$, and $\phi_{xy}=u'_y$, we obtain:
\begin{gather}\label{eq:eqn4}
    \left[a^2 - \left(U_\infty + u'\right)^2\right]\frac{\partial u'}{\partial x} + \left[a^2 - v'^2\right]\frac{\partial v'}{\partial y} - 2\left(U_\infty + u'\right)v'\frac{\partial u'}{\partial y} = 0.
\end{gather}
Recall that:
\begin{gather}\label{eq:eqn5}
    \frac{a_\infty^2}{\gamma - 1} + \frac{U_\infty^2}{2} = \frac{a^2}{\gamma - 1} + \frac{\left(U_\infty + u'\right)^2 + v'^2}{2}.
\end{gather}
By substituting this into equation~\eqref{eq:eqn4}, we get:
\begin{equation}\label{eq:eqn:5}
\begin{aligned}
    \left(1 - M^2_\infty\right)\frac{\partial u'}{\partial x} + \frac{\partial v'}{\partial y} &= M^2_\infty\left[\left(\gamma + 1\right)\frac{u'}{U_\infty}+ \frac{\gamma + 1}{2}\frac{u'^2}{U^2_\infty}+\frac{\gamma - 1}{2}\frac{v'^2}{U^2_\infty}\right]\frac{\partial u'}{\partial x}\\
    &+ M^2_\infty\left[\left(\gamma - 1\right)\frac{u'}{U_\infty} + \frac{\gamma + 1}{2}\frac{v'^2}{U^2_\infty}+\frac{\gamma - 1}{2}\frac{u'^2}{U^2_\infty}\right]\frac{\partial v'}{\partial y}\\
    &+ M^2_\infty\left[\frac{v'}{U_\infty}\left(1 + \frac{u'}{U_\infty}\right)\left(\frac{\partial u'}{\partial y} + \frac{\partial v'}{\partial x}\right)\right].
\end{aligned}
\end{equation}
With the small-disturbance assumption, typically valid for slender bodies at small angles of attack, we have $\frac{u'}{U_\infty}, \frac{v'}{U_\infty} \ll 1$. This implies that the right-hand side of equation~\eqref{eq:eqn:5} can be neglected, so the linearized perturbation velocity potential equation becomes:
\begin{gather}\label{eq:linear}
    \left(1 - M^2_\infty\right)\frac{\partial^2\phi}{\partial x^2} + \frac{\partial^2\phi}{\partial y^2} = 0.
\end{gather}

\subsection{Boundary Conditions}\label{sec:BC}
The velocity potential equation should satisfy two sets of boundary conditions: one at infinity and one at the body surface. At infinity, as the disturbance from the body vanishes, the flow approaches a uniform free stream, where $u \to U_\infty = 55 m/s$ and $v \to 0$.

At the body surface, the flow-tangency boundary condition, $\textbf{V}\cdot\textbf{n}=0$ ($\textbf{n}$ denotes the unit vector normal to the body surface), requires the velocity vector $\textbf{V}$ to be tangent to the body surface. In terms of the velocity components, this means that the ratio $v/u$ equals the slope of the surface tangent, i.e., $v/u=dy/dx$ on the body surface.

The elliptical geometry investigated in this paper is centered at the origin of a Cartesian coordinate system. We therefore define the body shape as:
\begin{gather}\label{eq:shape}
    x^2 + \frac{y^2}{b^2} = 1,
\end{gather}
where $b$ is an arbitrary constant between $0~\mathrm{m}$ and $1~\mathrm{m}$. The body shape is shown in Figure~\ref{fig:frame}.
To obtain the signed tangent slope, we treat $y$ in equation~\eqref{eq:shape} as an implicit function of $x$ and differentiate equation~\eqref{eq:shape}:
\begin{gather}\label{eq:deri}
    2x + \frac{2y}{b^2}\frac{dy}{dx} = 0.
\end{gather}
Rearranging equation~\eqref{eq:deri} and substituting $(x_0, y_0)$ into the equation, we obtain:
\begin{gather}\label{eq:theta0}
    \frac{dy}{dx} = -b^2\frac{x_0}{y_0}.
\end{gather}

For this elliptical geometry, we introduce the standard parametric angle $\theta\in[0, 2\pi)$ such that any point $(x_0, y_0)$ on the elliptical geometry can be written as $x_0=\cos\theta$, $y_0=b\sin\theta$. Substituting this into equation~\eqref{eq:theta0} and applying the tangency condition $\Phi_y/\Phi_x = dy/dx$ on the surface, we obtain the flow-tangency condition:
\begin{gather}\label{eq:flow-tangency}
    \frac{\Phi_y}{\Phi_x} = -b\frac{\cos\theta}{\sin\theta}.
\end{gather}
In the special case where $b = 1$, the geometry becomes a circular cylinder, and the flow tangency condition simplifies to:
\begin{gather}\label{eq:flow-tangency-circle}
    \frac{\Phi_y}{\Phi_x} = -\frac{\cos\theta}{\sin\theta}.
\end{gather}

\section{Methods}\label{sec:methods}

\subsection{Physics-Informed Neural Networks}\label{sec:PINNs}

Physics-Informed Neural Networks (PINNs) are used to solve physics problems that involve complex, nonlinear partial differential equations, usually expressed as:
\begin{gather}\label{eq:NLE}
    u_t + N[u] = 0, x\in\Omega, t\in [0, T].
\end{gather}
In the equation, $u(x, t)$ is the solution, $N$ is the nonlinear differential operator, and $\Omega$ represents the computational domain in $\mathbb{R}^d$.

PINNs generally consist of three main components: a fully connected neural network, an activation function $\sigma(\cdot)$ that processes the inputs and outputs of the fully connected network, and physics-informed loss functions that are used to optimize the network by minimizing the loss to zero. This process is used to approximate the solution $u(x,t)$.

The neural network has an input layer, an output layer, and several hidden layers in between. The input layer receives the coordinates $(x,t)$ and the output layer provides the approximation of $u(x,t)$. In each hidden layer, the relationship between the input $x_i$ and the output $y_j$ is described by the following equation:
 \begin{gather}\label{eq:activation}
     y_j = \sum_{i}w_{ji}x_i + b_j,
 \end{gather}
 where $w_{ji}$ are the weights, and $b_j$ are biases, which are adjusted by minimizing the mean squared error (MSE) of the loss functions. The MSE is calculated as follows:
 \begin{gather}
     \mathrm{MSE} = \mathrm{MSE}_0 + \mathrm{MSE}_b + \mathrm{MSE}_f.
 \end{gather}
 Here, $\mathrm{MSE}_0$, $\mathrm{MSE}_b$, and $\mathrm{MSE}_f$ represent the errors from the initial conditions, boundary conditions, and governing equations, respectively. They are defined as follows:
 \begin{gather*}
     \mathrm{MSE}_0 = \frac{1}{N_0}\sum_{i=1}^{N_0}\left|g_0\right|^2, \qquad 
     \mathrm{MSE}_b = \frac{1}{N_b}\sum_{i=1}^{N_b}\left|g\right|^2, \qquad
     \mathrm{MSE}_f = \frac{1}{N_f}\sum_{i=1}^{N_f}\left|f\right|^2,
 \end{gather*}
 where:
 \begin{gather*}
     g_0 = u\left(t_{0}^i,x_{0}^i\right)-u_{0}^i, \qquad
     g = u\left(t_{b}^i,x_{b}^i\right)-u_{b}^i,  \qquad
     f = u_t\left(t_{f}^i,x_{f}^i\right)+N\left[u\left(t_{f}^i,x_{f}^i\right)\right],
 \end{gather*}
In these equations, $g_0$, $g$, and $f$ represent the differences between the predicted and actual values for the initial conditions, boundary conditions, and governing equations, respectively. The terms $\left(x_{0}^i, t_{0}^i\right)$, $\left(x_{b}^i, t_{b}^i\right)$, and $\left(x_{f}^i, t_{f}^i\right)$ denote the initial data, collocation points on the boundary, and collocation points used for solving the governing equations. The numbers $N_0$, $N_b$, and $N_f$ correspond to the number of data points for each of these components. Derivatives in the loss functions are calculated using automatic differentiation, which is now widely supported in many frameworks such as Jax, PyTorch and TensorFlow.

\subsection{Normalization}\label{sec:normalization}
In this paper, we set the freestream velocity $U_\infty=55~\mathrm{m/s}$, which results in the original velocity potential equation having a magnitude much larger than unity. To improve the numerical stability of the solution process, we normalize the variables so that their values are of order unity. Specifically, we normalize freestream velocity as $U_\infty' =U_\infty/100=0.55~\mathrm{m/s}$. This leads to the relationship between the original and normalized perturbation velocity potentials, $\phi = 100\phi'$.

Since $a_\infty=U_\infty/M_\infty$, we can also normalize $a_\infty$ as $a_\infty'=a_\infty/100$, so that $a=100a'$. Hence, the normalized nonlinear velocity potential equation becomes:
\begin{gather}
    \left[1-\frac{1}{a'^{2}}\left(\frac{\partial\phi'}{\partial x}+U_\infty'\right)^2\right]\frac{\partial^2\phi'}{\partial x^2} + \left[1 - \frac{1}{a'^{2}}\left(\frac{\partial\phi'}{\partial y}\right)^2\right]\frac{\partial^2\phi'}{\partial y^2} - \frac{2}{a'^{2}}\left(\frac{\partial\phi'}{\partial x}+U_\infty'\right)\left(\frac{\partial\phi'}{\partial y}\right)\frac{\partial^2\phi'}{\partial x\partial y} = 0.
\end{gather}
Using this normalized relation, the linearized velocity potential equation takes the normalized form:
\begin{gather}
    \left(1 - M^2_\infty\right)\frac{\partial^2\phi'}{\partial x^2} + \frac{\partial^2\phi'}{\partial y^2} = 0.
\end{gather}

Note that after normalization, the results will also be normalized. Hence, to obtain the correct results, the output must be multiplied by the scaling factor of 100. For simplicity, we drop the prime notation and reuse the same symbols for the normalized variables (e.g., we write $\phi$ instead of $\phi'$ unless otherwise stated).

\subsection{Cost function for solving subsonic flow}
\begin{figure}[t!]
	\centering
	\includegraphics[width=1\textwidth]{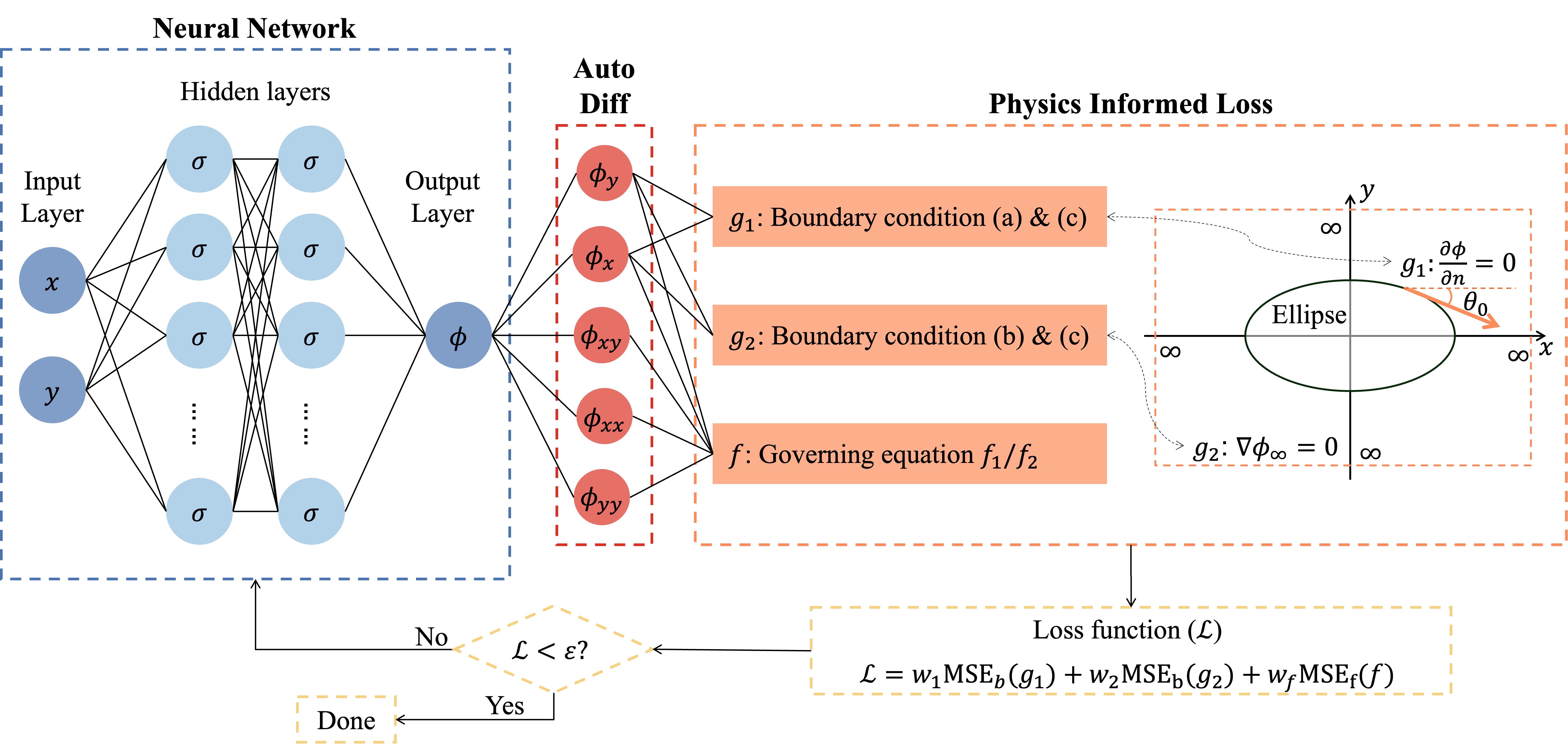} \vspace{-2em}
	\caption{\label{fig:frame} {\bf Schematic of PINN framework for solving subsonic flow.} The network consists of six hidden layers, each with 60 units, used to approximate the solution. The output $\phi$ is used to compute the loss functions through automatic differentiation, which are then employed to train the weights and biases of the activation function in the hidden layers. The right panel illustrates the imposed boundary conditions.}
\end{figure}

We now construct a PINN model to approximate the solution. The neural network consists of eight layers: one input layer with two inputs $x$ and $y$, one output layer with the approximation of $\phi(x, y)$, and six hidden layers, each with 60 units. A schematic of the PINN framework utilized to solve this problem is illustrated in Figure~\ref{fig:frame}.

The cost function is defined as a combination of two main terms: the boundary loss ($\mathrm{MSE}_b$) and the governing equation loss ($\mathrm{MSE}_f$). To define the cost functions, we first review the boundary conditions for this problem. These are:
\begin{enumerate}
    \item[(a)] Zero perturbation velocity in the far-field region;
    \item[(b)] Flow-tangency condition on the body surface;
    \item[(c)] Freestream velocity $U_\infty = 55~\mathrm{m/s}$.
\end{enumerate}
To satisfy boundary condition (a), we set $\phi_x$ and $\phi_y$ at infinity to zero. This leads the following boundary residual:
\begin{gather}\label{eq:loss1}
    g_1 = \{\phi_x(-\infty,y),\, \phi_x(\infty,y),\, \phi_y(x,-\infty),\, \phi_y(x,\infty)\}.
\end{gather}
As discussed in Section~\ref{sec:BC}, the flow tangency condition is given by equation~\eqref{eq:flow-tangency}. For the perturbation potential formulation, the flow tangency condition on the body surface is expressed as $\phi_y/(\phi_x+U_\infty)=-b\frac{\cos\theta}{\sin\theta}$. However, defining the loss function directly as $g_2= \phi_y/(\phi_x+U_\infty) + b\frac{\cos\theta}{\sin\theta}$ leads to singularities at $\theta =0, \pi$, and $2\pi$. To avoid this, we multiply both sides by $\sin\theta(\phi_x+U_\infty)$, and obtain the equivalent second boundary residual,
\begin{gather}\label{eq:loss2}
    g_2 = \phi_y\sin\theta + b\cos\theta(\phi_x+U_\infty).
\end{gather}
Thus, $g_1$ and $g_2$ define the boundary losses. For $g_1$, the collocation points are taken at infinity, while for $g_2$, they are taken on the body surface. Therefore, $g_1$ satisfies boundary condition (a), and $g_2$ satisfies boundary condition (b), while condition (c) is implicitly encoded in the decomposition $\Phi(x,y) = U_\infty x + \phi(x,y)$.

The equation residuals are defined as follows:
\begin{gather}
    f_1 =\left[1 - \frac{1}{a^2}\left(\frac{\partial\phi}{\partial x}+U_\infty\right)^2\right]\frac{\partial^2\phi}{\partial x^2} + \left[1 - \frac{1}{a^2}\left(\frac{\partial\phi}{\partial y}\right)^2\right]\frac{\partial^2\phi}{\partial y^2} - \frac{2}{a^2}\left(\frac{\partial\phi}{\partial x}+U_\infty\right)\left(\frac{\partial\phi}{\partial y}\right)\frac{\partial^2\phi}{\partial x\partial y},
\end{gather}
    
\begin{gather}
    f_2 = \left(1 - M^2_\infty\right)\frac{\partial^2\phi}{\partial x^2} + \frac{\partial^2\phi}{\partial y^2}.
\end{gather}
The residual $f_1$ is embedded in the nonlinear code, and $f_2$ in the linear code.

To control the relative importance of each loss term, we introduce weights $w_1$, $w_2$, and $w_f$ into loss terms. Typically, $w_1$ and $w_2$ are set to $1$, and $w_f$ is chosen to be less than $1$. Then the total loss is:
\begin{gather}
    \mathcal{L}= w_1\mathrm{MSE}_b(g_1) + w_2\mathrm{MSE}_b(g_2) + w_f\mathrm{MSE}_f(f).
\end{gather}

\section{Advanced settings for solving subsonic flow}\label{sec:advance}

\subsection{Coordinate transformation for infinite domains}\label{sec:coordinate}
\begin{figure}[t!]
	\centering
    \includegraphics[width=1\textwidth]{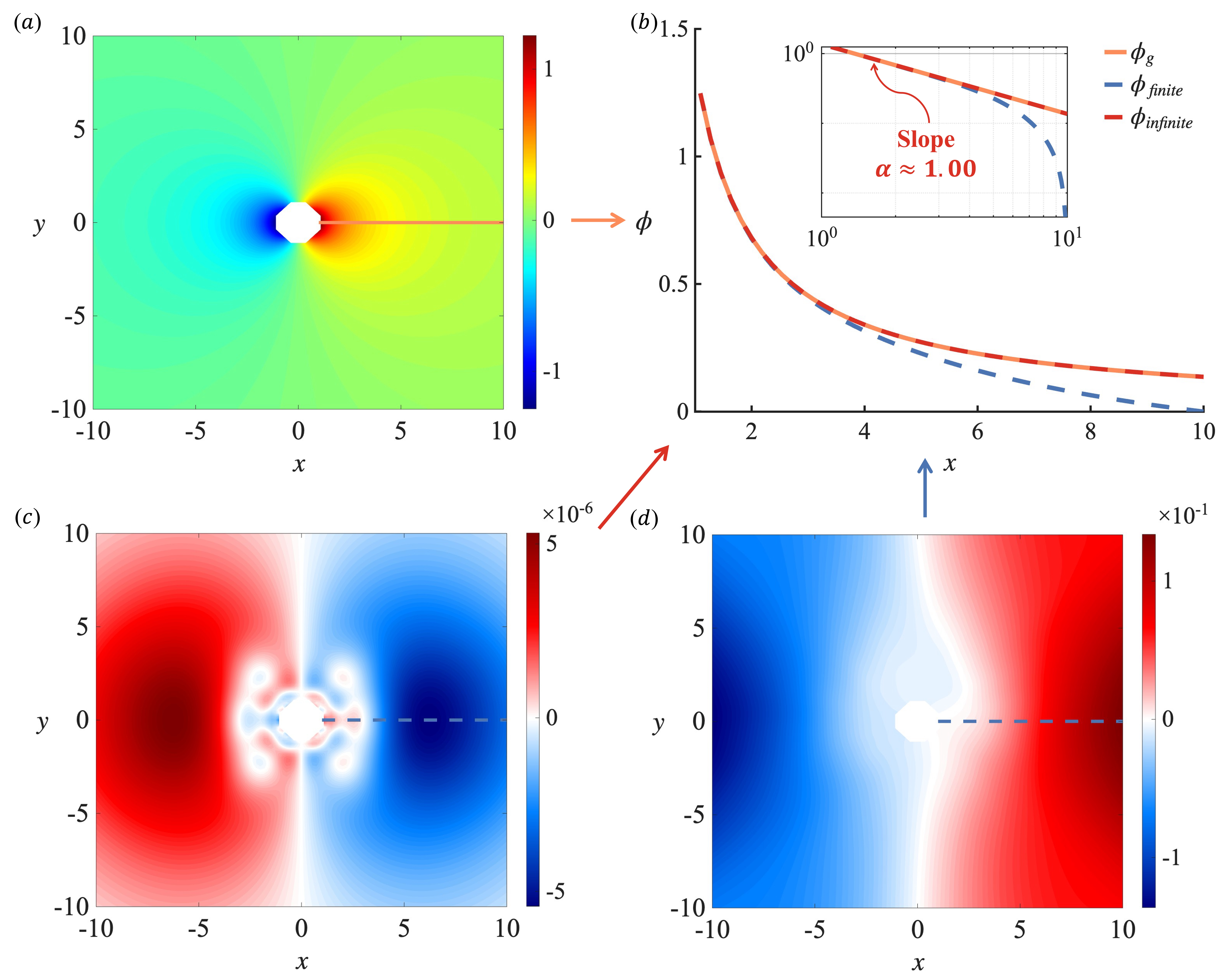}
    \caption{\label{fig:finite} {\bf Comparison of results between finite and infinite domains.} ($a$) Analytical solution \eqref{eq:incom_ana} in an infinite domain, visualized over the truncated region. ($b$) Comparison of solution profiles, accompanied by a log-log plot illustrating their asymptotic decay with distance. ($c$) Error distribution for the first-stage PINN solved on an infinite domain. ($d$) Error distribution for the first-stage PINN solved on a truncated finite domain.}
\end{figure}

When solving the problem in a truncated finite domain, a non-negligible truncation error is unavoidable. To quantify the effect of domain truncation, we consider a simplified benchmark that has a known analytical solution in an infinite domain: inviscid, irrotational, incompressible flow past a circular cylinder. The perturbation potential satisfies the Laplace equation:
\begin{gather}\label{eq:laplace}
    \frac{\partial^2\phi}{\partial x^2} + \frac{\partial^2\phi}{\partial y^2} = 0,
\end{gather}
which is equivalent to the polar form:
\begin{gather}
\label{eqn:incompressible}
\frac{1}{r}\frac{\partial}{\partial r}
\left(
r\frac{\partial \Phi}{\partial r}
\right)
+
\frac{1}{r^2}
\frac{\partial^2 \Phi}{\partial \theta^2}
=
0,
\end{gather}
where $(r,\theta)$ are polar coordinates. The analytical solution is given by:
\begin{gather}
\Phi(r,\theta)
=
U_\infty r \cos\theta
\left(
1 + \frac{R^2}{r^2}
\right).
\end{gather}
Writing $\Phi=U_\infty r\cos\theta+\phi$, the perturbation potential is:
\begin{gather}\label{eq:incom_ana}
\phi(r,\theta)
=
U_\infty \frac{R^2}{r} \cos\theta.
\end{gather}
Figure~\ref{fig:finite} compares the analytical infinite-domain solution with PINN solutions obtained in a truncated finite domain and in the compactified infinite domain using the coordinate transformation introduced later in this section. A clear discrepancy of order $\mathcal{O}(10^{-1})$ is observed, with errors concentrated predominantly in the far-field region. This shows that even if the near-body region is well resolved, the artificial outer boundary can still contaminate the solution unless the computational domain is extended sufficiently far.

\begin{figure}[t!]
	\centering
    \includegraphics[width=\textwidth]{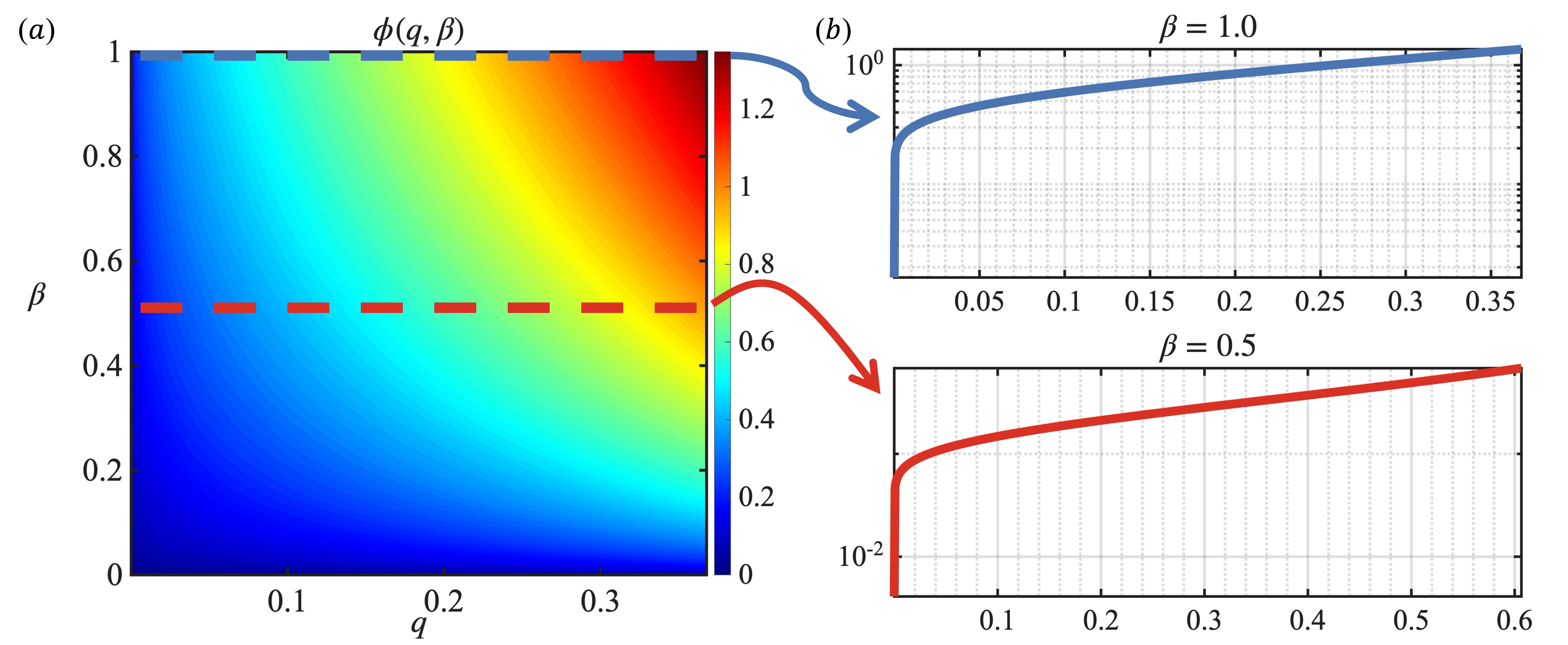}
    \caption{\label{fig:exp_transformation} {\bf Solution profile under exponential mapping ($q=e^{-r}$).} For incompressible flow, (a) shows $\phi(q,\beta)$ in the $(q,\beta)$ domain; (b) shows slices at $\beta=1$ and $\beta=0.5$, highlighting a sharp gradient near $q=0$.}
\end{figure}

Directly extending the domain to infinity in Cartesian coordinates is impractical for PINNs because sampling becomes inefficient and numerical conditioning can deteriorate. Instead, we need a mapping that compactifies the infinite physical domain into a finite computational domain. Many compactification mappings are possible. A common choice is $q=e^{-r}$ with $r=\sqrt{x^2+y^2}$ and $\beta=\frac{x^2}{x^2+y^2}$. After transforming the solution to $\phi(q,\beta)$, as illustrated in Figure~\ref{fig:exp_transformation}($a$), the infinite domain is indeed mapped into a finite interval and the decay appears smooth in the plot. However, Figures~\ref{fig:exp_transformation}($b$) show that $\phi$ develops a sharp gradient as $q\to0$. In particular, $\partial\phi/\partial q$ tends to infinity as $q\to0$, leading to a singularity at $q=0$ (i.e., $r\to\infty$) and making the problem difficult to solve numerically.

Thus, to design a valid compactification, we first examine the far-field behavior of the solution. For equation~\eqref{eq:laplace}, the analytical solution \eqref{eq:incom_ana} shows that the perturbation potential decays algebraically as $\phi \propto r^{-1}$, which motivates a power-law-based radial compactification. Figure~\ref{fig:finite}($b$) also confirms this decay numerically. In physical coordinates, $\phi$ changes rapidly near the body and transitions to a slowly varying far-field tail. A log-log plot of $|\phi|$ versus $r$ is close to a straight line with slope magnitude near one, indicating:
\begin{gather}
\phi \propto r^{-\alpha},
\qquad
\alpha \approx 1.
\end{gather}
Therefore, a coordinate transformation for the infinite domain should follow this asymptotic structure. In particular, it is desirable to introduce a transformed radial coordinate that scales linearly with the dominant far-field decay of $\phi$.

Motivated by this consideration, we introduce transformed coordinates $(q,\beta)$ defined by:
\begin{equation}
\begin{gathered}
q= (1+r^2)^{-\alpha/2}
= (1+x^2+y^2)^{-\alpha/2},\\
\beta= \frac{x^2}{r^2}
= \frac{x^2}{x^2+y^2},
\end{gathered}
\label{eq:qbeta}
\end{equation}
where $\alpha>0$ controls the radial scaling. This mapping transforms the infinite physical domain $(x,y)\in(-\infty,\infty)^2$ into the compact and normalized computational domain $(q,\beta)\in[0,1]\times[0,1]$. The coordinate $q$ captures the dominant radial variation, while $\beta$ represents the angular dependence. Note that this mapping also imposes even symmetry by construction, which can be problematic when the target solution has mixed symmetry. We address this issue in Section~\ref{sec:embedding}.

When $\alpha=1$, the transformed coordinate satisfies $q \sim r^{-1}$ in the far field, making $q$ linearly proportional to the leading-order perturbation potential. This choice improves the conditioning of the learning problem because the dominant decay is absorbed into the coordinate system rather than being learned by the network. More generally, $\alpha$ has a clear physical meaning: it represents the asymptotic decay exponent of the perturbation potential. If $\alpha$ is chosen to match the asymptotic behavior, the transformed solution varies smoothly over the compact domain. If $\alpha$ is chosen incorrectly, the mapping becomes misaligned with the solution structure. Specifically, a value that is too large over-compresses the far field near $q=0$, which amplifies gradients and worsens numerical conditioning, while a value that is too small does not capture the decay well and makes the far-field behavior harder to learn. In both cases, convergence deteriorates and the solution accuracy decreases.

Finally, we also tested an alternative reduced-order approach that solves directly for the perturbation velocity components $u'$ and $v'$ instead of the potential $\phi$. However, in our PINN experiments, this velocity-based formulation did not outperform the potential-based formulation as discussed in \ref{sec:velocity_eq}

\subsection{Physical embedding in neural networks}\label{sec:embedding}

Building on the previous coordinate transformation and physical insights, we further embed physical constraints, denoted by $q$, into the neural network. The prediction is formulated as:
\begin{equation}
\begin{gathered}\label{eq:hard_constraint}
\phi_{pred}(x,y)=q\cdot\mathcal{N}[q,\beta],\\
    \text{where}\quad q=(1+x^2+y^2)^{-1/2},\quad \beta=\frac{x^2}{x^2+y^2}, \quad (x,y)\in\Omega:=\mathbb{R}^2\setminus\mathcal{B},
\end{gathered}
\end{equation}
where $\mathcal{N}$ denotes the neural network and $\phi_{pred}$ is the model prediction. Here $\mathbb{R}^2$ denotes the unbounded physical plane; in practice, collocation points are sampled in $(q,\beta)\in(0,1]\times[0,1]$ and mapped back to the exterior flow domain $\Omega=\mathbb{R}^2\setminus\mathcal{B}$, where $\mathcal{B}$ is the solid body (for the ellipse, $\mathcal{B}=\{(x,y):x^2+y^2/b^2\le 1\}$). This formulation incorporates the asymptotic power-law decay into the network, allowing the neural network to focus only on learning the remaining linear expression and simplifying the training process. Additionally, the imposed decay constraint ensures that the solution approaches zero at infinity, consistent with the physical boundary condition.

Nevertheless, as the imposed decay forces the solution to approach zero at infinity, this restriction can cause the network to underemphasize the loss contribution from the far-field region, causing the PINN to neglect this part of the domain. To address this, additional techniques are needed to amplify the residuals in that region, which will be discussed in the following sections.

Additionally, the target solution is not symmetric in both coordinate directions. Specifically, the perturbation velocity potential $\phi(x,y)$ is odd in $x$ and even in $y$. To integrate this symmetry into the network, we impose the following constraint, and modify the model to:
\begin{gather}\label{eq:symmetric_constraint}
\phi_{pred}(x,y)=\frac{x}{\sqrt{x^2+y^2}} \cdot q\cdot\mathcal{N}[q,\beta].
\end{gather}
This formulation ensures that the model respects the physical symmetry of the flow field while retaining flexibility through the learnable neural representation.

\subsection{Collocation point arrangement}\label{sec:collocation}

PINNs solve partial differential equations by selecting specific collocation points within the domain where the solutions are enforced to satisfy the governing equations. The proper selection of these points is crucial for achieving accurate and efficient solutions, especially in complex domains. In this study, we used the Latin Hypercube Sampling (LHS) method to randomly generate points across the domain space.

However, the irregular boundary, near which flow changes sharply, causes difficulties in solving the governing equations. While transforming the irregular domain into a regular one can sometimes help, this approach has limitations, as it requires a different transformation for each specific case. Additionally, the infinite mapping discussed earlier introduces another challenge, which maps smaller segments to relatively larger regions. For instance, a small interval such as $[0,10]$ is mapped to $[0.1,1]$, while the infinite domain $[10,+\infty]$ is compressed into $[0,0.1]$, causing the PINN to give insufficient attention to the far-field region. This imbalance, where the small regions receive more focus than larger ones, impacts the accuracy of the solution.

To address these challenges, two additional strategies are employed: additional boundary points and adaptive sampling. The first strategy involves placing extra collocation points at boundaries, sharply changing regions or the infinite region based on the performance of the training. These points can be evenly distributed or placed according to the behavior of the function. The second strategy, adaptive sampling, dynamically adjusts the density of collocation points based on error distribution during training. Specifically, the equation error distribution, $F$, is calculated across the domain and normalized to generate a normalized error distribution, $F_{nm}$, during a specific iteration of the training. However, this approach can lead to the concentration of collocation points at certain locations, causing unbalanced training and potential low-frequency and biased results. To mitigate this, a sensitive index, $\Delta f$, is introduced to reduce the difference and sensitivity of the error distribution. Typically, $\Delta f$ is selected between $0.1$ an $0.5$.
 \begin{gather}\label{eq:adaptive_normalize}
     F=f^2,\\
     F_{nm}=\frac{F}{\mathrm{MSE}_f(f)}+\Delta f.
 \end{gather}

To reduce noise and increase the efficiency of sampling, a Gaussian filter is used to smooth the normalized error distribution:
 \begin{gather}\label{eq:adaptive_smooth}
     F_s=\mathcal{G}(F_{nm},\sigma,w).
 \end{gather}
Here, $\mathcal{G}$ is the Gaussian smoothing operator, with $\sigma$ and $w$ representing the standard deviation and width of the filter, respectively. The error at each node is combined with the errors from its surrounding $w$ neighboring nodes, effectively smoothing the distribution. For irregular domains, the process involves two critical steps to ensure robust Gaussian smoothing. First, before smoothing, a semi-mask is applied to $F_{nm}$ to account for the regions outside the domain by extending $w/2$ grids beyond the irregular boundary. This ensures balanced Gaussian smoothing while slightly increasing the error probability near the irregular boundary. This is particularly important because the irregular boundary presents a challenge for training. Secondly, after smoothing, the result $F_s$ is fully masked within the body region, ensuring that no points are generated outside the domain.

Once the error distribution is smoothed, the cumulative distribution function, $B$, is calculated. A random number, $c$, is then generated between $[0,B_{max}]$ to select grid points based on the relative probabilities. Finally, sampling points are fine-tuned by introducing small random offsets to the selected grid points, ensuring that the sampling points are not strictly confined to the grid points.

These approaches ensure a higher density of sampling points in regions with higher errors, enhancing the overall efficiency and accuracy of the model in complex areas.

\subsection{Multi-stage PINNs for irregular boundary condition}\label{sec:MSPINN}

Wang and Lai \cite{WANG2024112865} proposed a multi-stage neural network framework, which resolves the precision bottleneck by introducing multiple stages of networks. In each stage, a new neural network is introduced and optimized to learn the residual from the previous stage. A frequency spectrum-based method is used for the transition, and the process iterates to achieve machine-precision results with an exponential decay rate. However, the frequency spectrum-based stage transition method is not directly applicable on an irregular domain, since a standard DFT requires values on a full uniform rectangular grid. Additionally, coordinate transformation makes the training more difficult, as there is unequal accuracy between the inner finite region and the outer infinite region.

\subsubsection{Frequency spectrum-based transition for irregular domain}

To begin, we review the original spectrum-based error relation used for stage transition. Once the previous stage is properly trained and its boundary condition is close to zero, the following equation holds:
\begin{equation} \label{eq:ms_frequency}
\begin{gathered}
    \epsilon_1 = \frac{\epsilon_{r1}}{[2\pi f_d^{(r)}]^m \cdot \epsilon_\beta}, \\
    \mathrm{with} \ \epsilon_1 = \mathrm{RMS}(e_1(x,u_0)), \ 
    \epsilon_\beta = \mathrm{RMS}(\beta_m), \ 
    \epsilon_{r1} = \mathrm{RMS}(r_1(x,u_0)).
\end{gathered}
\end{equation}
Here, $\epsilon_{1}$, $\epsilon_\beta$ and $\epsilon_{r1}$ represent the magnitudes of errors $e_1(x,u_0)$ (the error between the exact solution and the predicted solution), $\beta_m$ (the coefficient of the highest-order derivative), and $r_1(x,u_0)$ (the equation residual of the previous stage), respectively. $f_d^{(r)}$ denotes the dominant frequency of the equation residual $r_1(x,u_0)$, and $m$ refers to the highest order of the derivative of $u$. $\mathrm{RMS}(\cdot)$ denotes the root mean square error (RMS) of a quantity evaluated on the sampled points.

For problems where the residual is available on a full, uniformly sampled rectangular grid, the dominant frequency can be determined using the Discrete Fourier Transform (DFT). However, for PDEs over an irregular domain, a standard DFT cannot be applied directly. While coordinate transformations are often used to address this issue, they limit generalization, particularly when dealing with complex boundaries. Another option, domain decomposition, could also be used to solve the problem, but with less accuracy. As discussed in Section~\ref{sec:collocation}, the infinite region exhibits lower frequencies than the inner region, leading to potential frequency differences across the domain that limit the effectiveness of domain decomposition methods.

To address this issue, we leverage the nature of the equation residual, $r_1(x,u_0)$, which is theoretically zero and closely approximates zero. By assigning zero to the region outside the irregular domain, we can make the DFT work effectively. This method has two main advantages: it is highly generalizable, making it convenient and effective for handling irregular boundaries, and it does not introduce significant errors due to the near-zero residual. However, potential ambiguities, such as the Gibbs phenomenon and spectral leakage, may appear in the spectrum due to the arbitrary masking of the error. These issues can be easily detected without concern, and further discussion is provided in Section~\ref{sec:spectral artifact}

\subsubsection{MS-PINN setup, hyperparameters, and collocation point}

The key hyperparameters of the multi-stage network are the magnitude prefactor (the normalization factor for the new stage), $\varepsilon_1$, the modified scale factor (the layer scale of the first layer of the network, which determines the frequency of the solution), $\hat{\kappa}_1$, and the equation weight $w_f$, which are defined as:
\begin{gather}\label{eq:ms_setup}
    \varepsilon_1=\epsilon_1,\ \hat{\kappa}_1=[2\pi f_d^{(r)}]^m\cdot\epsilon_1,\ w_{f1}=\frac{w_{f0}}{[2\pi f_d^{(r)}]^m},
\end{gather}
where $w_{f0}$ refers to the equation weight of the previous stage and $\hat{\kappa}_1$ can be slightly increased to capture higher frequencies of results of the new stage.

A common misconception about this relation is that higher frequencies lead to higher accuracy. While increasing the modified scale factor can raise the frequency of the solution, the accuracy of the higher-stage network is governed by the homogeneity of both the frequency and magnitude of the previous stage. Large variations in frequency or significant differences in the magnitude of target functions and residual terms can lead to training failure, as the network may get stuck in local minima. This happens because the optimization landscape becomes non-convex, and the network tends to prioritize fitting low-frequency and high-magnitude components while neglecting others.

Furthermore, to achieve machine-precision solutions, the collocation points must be carefully designed. In addition to the strategy discussed in Section~\ref{sec:collocation}, the strategy of adding boundary points may vary between the first and second stages when solving equations defined over an infinite domain after the coordinate transformation. In the first stage, the learning difficulty across both the infinite region and the inner region around the body is relatively balanced. Thus, concentrating boundary points or even omitting them may still yield good results. However, in the second stage, as the overall error decreases and the model approaches higher precision, small variations in the infinite region-previously negligible-become relatively significant. These variations dominate the local error and attract more focus from the network. Unfortunately, due to the highly compressed nature of the transformed coordinates, this region occupies a narrow interval (e.g., $\mathcal{O}(+\infty)$ mapping to $\mathcal{O}(0.1)$) and contains relatively few collocation points. As a result, the network struggles to fit the function accurately in this area. As a result, training in the infinite region becomes more challenging in the second stage, making the use of additional points near the infinite boundary essential.

\subsubsection{Performance validation and final adjustments}

\begin{figure}[t!]
	\centering
	\includegraphics[width=1\textwidth]{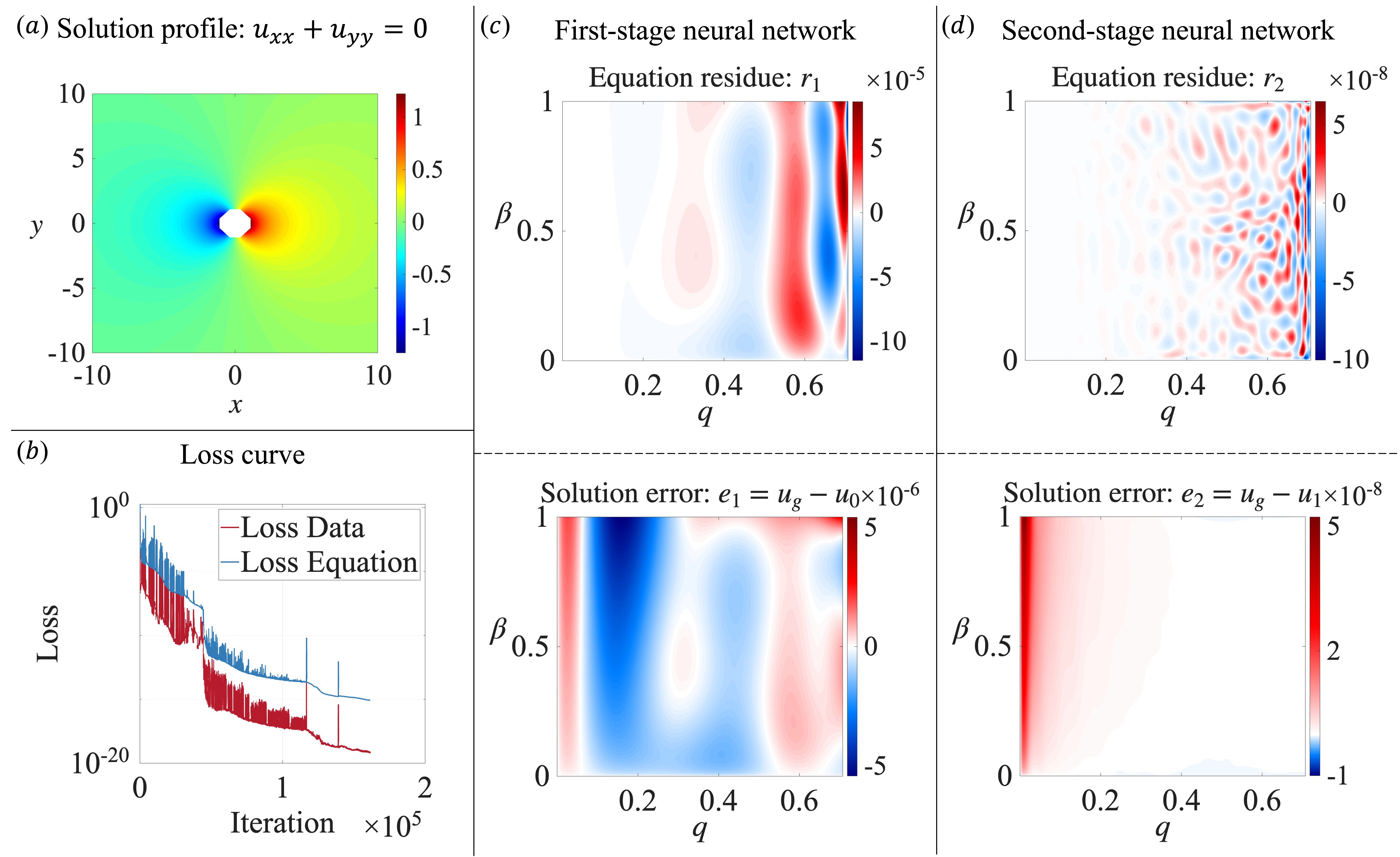}
	\caption{\label{fig:laplace} {\bf MS-PINN solution for incompressible flow.} ($a$) Predicted potential $\phi$ for the Laplace equation~\eqref{eq:laplace}. ($b$) Loss curve over multi-stage training. ($c$) Equation residual and solution error after the first-stage training. ($d$) Corresponding results after the second-stage training.}
\end{figure}

To validate the performance of MS-PINN, we first apply it to solve the 2D incompressible and inviscid flow around a circular cylinder, where the perturbation potential satisfies the Laplace equation~\eqref{eq:laplace}. Figure~\ref{fig:laplace} shows the results, evaluated in terms of both the equation residual and the solution error. The results demonstrate that MS-PINN attains machine-precision accuracy, with RMS on the order of $\mathcal{O}(10^{-8})$, highlighting its superior capability in solving PDEs.

Note that the relationship between the equation error and the solution error does not strictly follow equation~\eqref{eq:ms_frequency}. This discrepancy arises primarily from insufficient training in the infinite domain, even with additional collocation points. Since the hard constraint~\eqref{eq:hard_constraint} is imposed at infinity to enforce zero values, the error in the infinite region is naturally reduced to zero, which misleads the optimizer.

To improve accuracy and bring the results closer to equation~\eqref{eq:ms_frequency}, we modify the governing equation loss by dividing it by $q^{\gamma}$, a weight adjustment factor that amplify the error in the infinite region. The modified equation loss is:
\begin{gather}\label{eq:modify_loss}
     MSE_f = \frac{1}{N_f}\sum_{i=1}^{N_f}\left|f/q^{\gamma}\right|^2,
\end{gather}
where $\gamma$ is the magnification factor, typically ranging from $0.5$ to $3$. In practice, the magnification factor is usually higher during the first-stage training and can be adjusted until the training accuracy deteriorates. The failure of the $q^{\gamma}$ setup can generally be identified by several characteristics: the equation loss does not decrease, remains abnormally high, or fails to converge; An excessive number of adaptive collocation points concentrates in the infinite region; the resulting equation error is unsatisfactory or anomalously large; and the error exhibits low-frequency features with abnormalities in its frequency spectrum. Any of these signs typically indicates that $\gamma$ is set too high.

A common phenomenon after such an operation might appear to contradict our earlier statement that the frequency in the first-stage training should remain homogeneous. However, when examining the equation error and the frequency spectrum of $f/q^{\gamma}$ instead of $f$ alone, we find that the frequency distribution remains consistent. This observation confirms the success of the first-stage training and demonstrates that the adjustment does not conflict with our previous proposition.

Although this strategy improves the solution, it is important to note that even without this modification, MS-PINN already achieves satisfactory accuracy of around $\mathcal{O}(10^{-7})$.

\section{Results}\label{sec:results}

In this section, we present results in order of increasing complexity. After validating the method on the Laplace equation in the previous section, we first apply it to a more complex linear formulation. We then extend the model to the nonlinear problem and compare the linear and nonlinear solutions to quantify the impact of nonlinearity. Finally, we test the method on more complex geometries.

\subsection{Linear and nonlinear compressible subsonic flow}\label{sec:linear_nonlinear}

In Section 4, we introduced and validated our method using the Laplace equation~\eqref{eq:laplace}, which demonstrates the robustness and accuracy of the model. We now extend the same approach to a more challenging linear compressible equation~\eqref{eq:linear}. This linear equation has a similar form to equation~\eqref{eq:laplace}, but includes additional coefficients, so it generally does not admit a simple closed-form solution for this boundary-value problem. Because the overall form is still similar, we can use the same MS-PINN setup as before.

\begin{figure}[t!]
	\centering
	\includegraphics[width=1\textwidth]{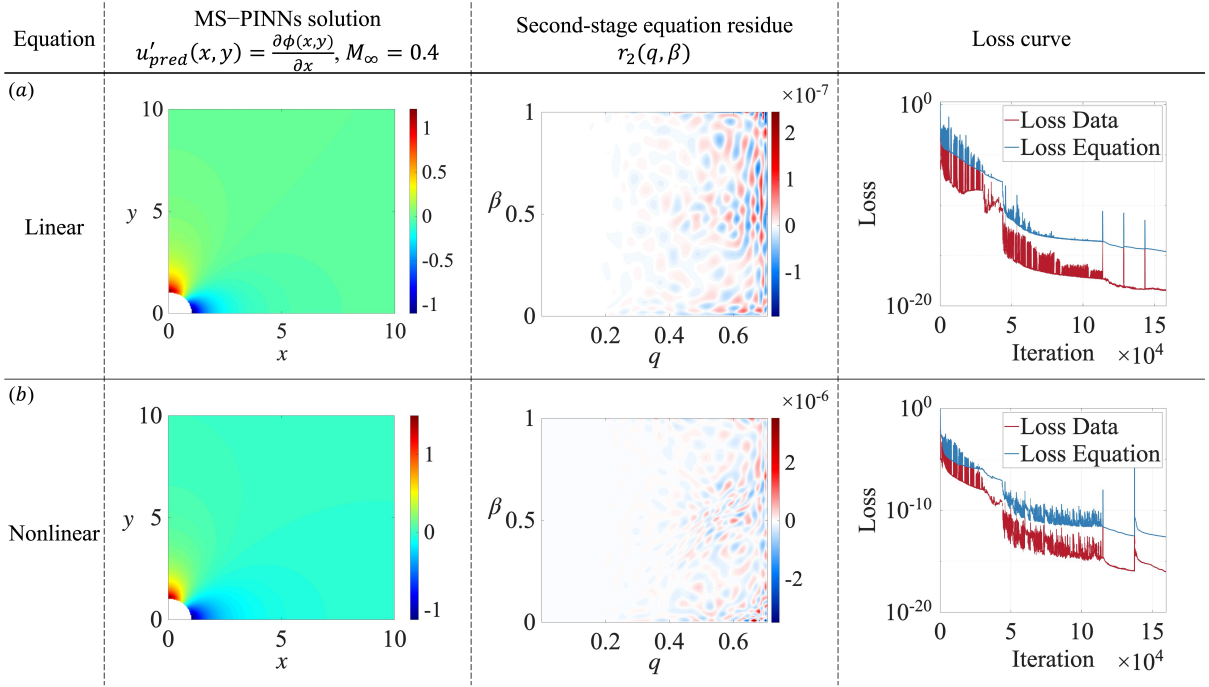}
	\caption{\label{fig:linear_nonlinear} {\bf MS-PINN solutions for compressible flow}. ($a$) Solution of the linear equation~\eqref{eq:linear} at $M_{\infty}=0.4$, shown via $u'=\partial\phi/\partial x$, together with the second-stage equation residual, $r_2(x,y)$, over the infinite domain and the corresponding loss curves. ($b$) Corresponding results for the nonlinear equation~\eqref{eq:eqn1_pert}.}
\end{figure}

Figure~\ref{fig:linear_nonlinear}($a$) presents the MS-PINN results for the linear equation at Mach number $M_\infty = 0.4$. We plot the solution over a finite region, report the second-stage equation residual over the infinite domain, and include the corresponding loss curves. The results reach machine-precision accuracy.

We further apply the method to the nonlinear equation~\eqref{eq:eqn1_pert}, which has long been difficult to solve accurately and efficiently. As illustrated in Figure~\ref{fig:linear_nonlinear}($b$), the results demonstrate that MS-PINN can robustly achieve an RMS error of $\mathcal{O}(10^{-7})$. The method also remains computationally efficient, showing that it can directly capture nonlinear effects that are missed by linear models.

\begin{figure}[t!]
	\centering
	\includegraphics[width=1\textwidth]{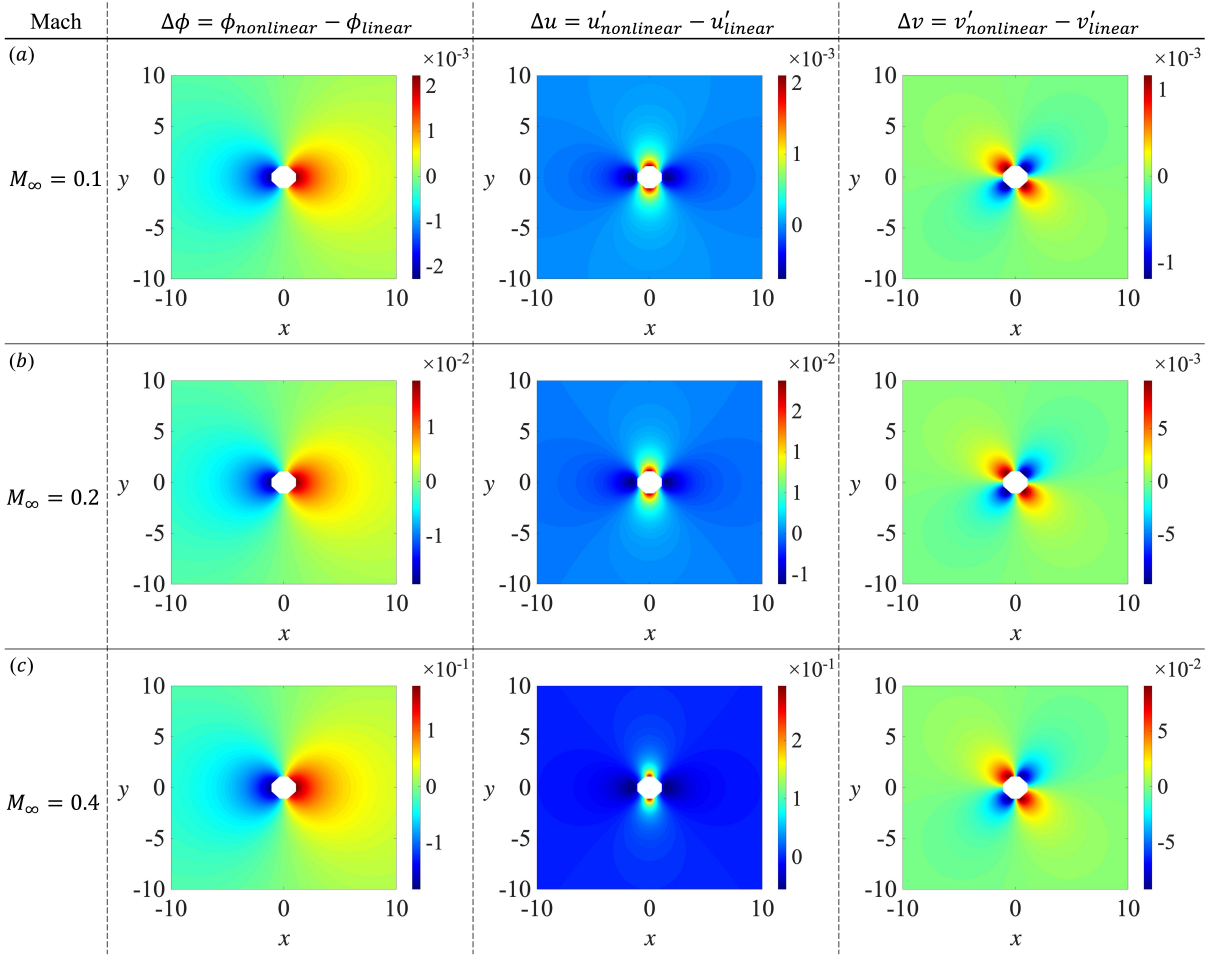}
	\caption{\label{fig:linear_vs_nonlinear} {\bf Comparison between linear and nonlinear equations}. Spatial distributions of the differences in velocity potential, $\Delta\phi$, streamwise velocity $\Delta u$, and transverse velocity $\Delta v$ for freestream Mach numbers of ($a$) $M_\infty=0.1$, ($b$) $M_\infty=0.2$, and ($c$) $M_\infty=0.4$, illustrated over a finite domain.}
\end{figure}

With this level of accuracy, MS-PINN provides a sufficiently accurate solution to resolve the differences between the linear and nonlinear models. The observed discrepancies are on the order of $\mathcal{O}(10^{-3}\sim10^{-1})$, which are several orders of magnitude larger than the numerical error. Therefore, these differences reflect genuine nonlinear effects rather than errors caused by limited solver accuracy. Figure~\ref{fig:linear_vs_nonlinear} illustrates the differences between the linear and nonlinear equations at increasing Mach numbers, plotted over the same finite region for consistency. As the Mach number rises from 0.1 to 0.4, the discrepancy grows rapidly: it is about $\mathcal{O}(10^{-3})$ near the body at $M_\infty = 0.1$, increases to $\mathcal{O}(10^{-2})$ at $M_\infty = 0.2$, and reaches $\mathcal{O}(10^{-1})$ at $M_\infty = 0.4$. In relative terms, the difference around the body increases from $0.6\%$ to about $13\%$, showing a nearly exponential increase in discrepancy with Mach number. The largest deviations are concentrated near the body surface: compared with the nonlinear solution, the linear model predicts a higher potential on the windward side and a lower potential in the wake, reflecting a systematic overestimation of the velocity potential upstream and an underestimation downstream.

\subsection{Extension to Other Geometries}
\label{sec:geometry}

We further extend the method to more complex geometries. As an example, we consider an elliptical geometry defined by $x^2 + 4y^2 = 1$. From the boundary residual in equation~\eqref{eq:loss2}, when $b=0.5$, the boundary residual becomes:
\begin{gather}\label{eq:ellipse_loss}
    g_2' = \phi_y\sin\theta + 0.5\cos\theta(\phi_x+U_\infty).
\end{gather}

We consider two ways to handle this geometry. The first approach involves transforming the coordinate system by setting $y' = 2y$. The transformed variables are then:
\begin{gather}\label{eq:ellipse_coordinate}
    q=(1+x^2+y'^2)^{-1/2}=\left(1+x^2+(2y)^2\right)^{-1/2},\\
    \beta =\frac{x^2}{x^2+(2y)^2}.
\end{gather}
With this scaling, the transformed coordinates remain within a rectangular region, which simplifies the computational process and maintains numerical stability. The second, more general approach retains the original coordinate transformation while allowing the transformed domain to remain irregular. This avoids geometry-specific scaling and can be applied directly to arbitrary shapes.

\begin{figure}[t!]
	\centering
	\includegraphics[width=1\textwidth]{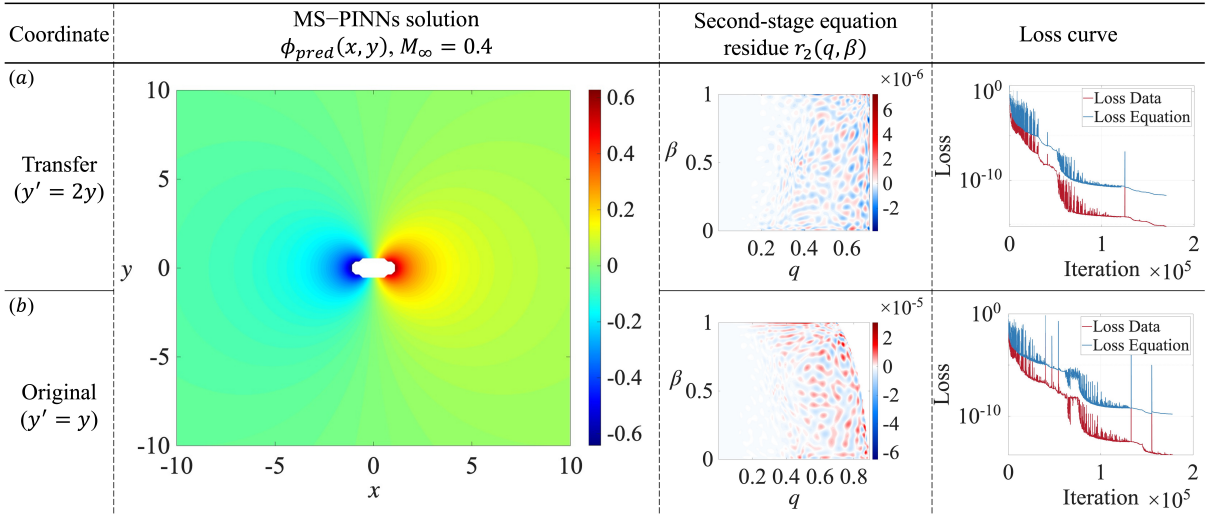}
	\caption{\label{fig:ellipse} {\bf MS-PINN solutions over an ellipse}. ($a$) Solution profile for the nonlinear equation~\eqref{eq:eqn1_pert} of compressible subsonic flow over an ellipse at $M_{\infty}=0.4$, obtained using MS-PINN with the scaled transformation $y'=2y$, including the predicted potential $\phi_{pred}(x,y)$, the second-stage equation residual $r_2(x,y)$, and the loss curves. ($b$) Corresponding results using the original transformation $y'=y$.}
\end{figure}

Figure~\ref{fig:ellipse} shows that the method can be applied accurately using either approach. Both methods produce accurate solutions, with only small differences in the final results. The scaled case ($y'=2y$) reaches an accuracy of about $\mathcal{O}(10^{-7})$, while the unscaled case ($y' = y$) reaches about $\mathcal{O}(10^{-6})$. Overall, these results demonstrate that the proposed model remains accurate and robust when extended to complex geometries.

\section{Conclusion}\label{sec:conclusion}

This study presents a new computational framework for solving nonlinear subsonic compressible flow in an infinite domain using Physics-Informed Neural Networks. The difficulty of this problem stems primarily from two aspects: the nonlinearity of the governing equations and the unbounded nature of the domain. Guided by the asymptotic behavior of the linearized equation in the far-field region, we successfully introduced a coordinate transformation that maps the unbounded physical domain to a normalized computational space, where the solution profile maintains high regularity. By employing this transformed coordinate system as the network input, we effectively solve the equation over an infinite domain, thereby eliminating the need for domain truncation and the noticeable errors associated with it.

To enhance training efficiency and accuracy, we further proposed a solution ansatz for the perturbation potential $\phi$, constructed by multiplying the network output by the solution's asymptotic decay. Consequently, the solution automatically satisfies the boundary condition at infinity without the need for soft constraints within the loss function. Our tests confirm that this hard-constraint formulation accelerates convergence and prevents the optimization from stagnating in local minima. Furthermore, by integrating this architecture with a multi-stage training strategy (MS-PINN), we successfully reduced equation residuals to the level of machine precision. We note that this framework is effective not only for solving the linearized subsonic flow equation but can also be directly applied to solve the original nonlinear equations.

Leveraging this high-precision framework, we conducted a comparative analysis between the linearized perturbation velocity potential equation and the full nonlinear formulation. We demonstrated that the error inherent in the linearized approximation increases nonlinearly with the Mach number, underscoring the necessity of full nonlinear solvers for higher-speed subsonic regimes. Additionally, we quantified the accuracy improvements gained by solving in an infinite domain versus a truncated domain. Finally, the method’s generalizability was validated through application to elliptical geometries. Future work will focus on extending this methodology to complex, asymmetric body shapes and higher-dimensional aerodynamic problems in infinite domains.

\setcounter{section}{0}
\renewcommand\thesection{Appendix \Alph{section}.}
\renewcommand\theequation{\Alph{section}.\arabic{equation}}

\section{Reduced order of PDE}\label{sec:velocity_eq}

We note that the gradient of the perturbation potential, $\nabla\phi$, gives the perturbation velocity. Since $\nabla\phi\propto r^{-2}$, the perturbation velocity decays to zero in the far-field region. Compared with the potential formulation, the governing equations in the velocity formulation only require first-order derivatives of $u'$ and $v'$, which may reduce the training difficulty. We therefore rewrite the governing equations in terms of $u'$ and $v'$. Using the potential-velocity relations:
\begin{gather}
    \frac{\partial\Phi}{\partial x} = \frac{\partial\phi}{\partial x} + U_\infty = u' + U_\infty,\\
    \frac{\partial\Phi}{\partial y} = \frac{\partial\phi}{\partial y} = v',
\end{gather}
the nonlinear perturbation velocity potential equation can be rewritten as:
\begin{gather}
    \left[1 - \frac{1}{a^2}\left(u' + U_\infty\right)^2\right]\frac{\partial u'}{\partial x} + \left[1 - \frac{1}{a^2}v'^2\right]\frac{\partial v'}{\partial y} - \frac{2}{a^2}\left(u' + U_\infty\right)v'\frac{\partial u'}{\partial y} = 0 \label{eq:nonlinear_velocity_GE},
\end{gather}
together with the irrotationality constraint,
\begin{gather}
    \frac{\partial u'}{\partial y} - \frac{\partial v'}{\partial x} = 0.
    \label{eq:irrotationality_GE}
\end{gather}
For the linearized case, the equation becomes:
\begin{gather}
    \left(1 - M^2_\infty\right)\frac{\partial u'}{\partial x} + \frac{\partial v'}{\partial y} = 0. \label{eq:linear_velocity_GE}
\end{gather}

However, solving the coupled velocity formulation \eqref{eq:nonlinear_velocity_GE} or \eqref{eq:linear_velocity_GE} together with the irrotationality constraint \eqref{eq:irrotationality_GE} is more challenging for PINNs. In the potential formulation, the two velocity components are tightly coupled through a single scalar field $\phi$, so the coupling is naturally enforced. In contrast, in the velocity formulation, the coupling relies only on \eqref{eq:irrotationality_GE}, which provides a much weaker constraint during PINN optimization. In our experiments, the PINN solution often becomes nearly constant, with high residuals remaining in the far-field region. This issue becomes worse after the infinite-domain coordinate transformation. The region where the error is large (the far-field region) is further compressed, and the loss is averaged across the domain. Consequently, the high-error far-field region contributes little to the total loss and is more easily ignored by the optimizer, making the solution harder to train.

To alleviate this, we introduce a hard decay constraint for the velocity formulation together with the transformed coordinates:
\begin{equation}
\begin{gathered}\label{eq:uv_hard_constraint}
    u_{pred}(x,y)=q\cdot\mathcal{N}[q,\beta],\\
    \text{where}\quad q=\frac{1}{1+x^2+y^2},\quad \beta=\frac{x^2}{x^2+y^2}, \quad (x,y)\in\Omega:=\mathbb{R}^2\setminus\mathcal{B}.
\end{gathered}
\end{equation}
Here, $\mathcal{N}$ denotes the neural network, and $u_{pred}(x,y)$ is the model output. As before, $\mathbb{R}^2$ denotes the unbounded plane and $\mathcal{B}$ denotes the solid body. We choose $q=(1+x^2+y^2)^{-1}$ because the perturbation velocity decays as $r^{-2}$ in the far-field region. Multiplying the network output by $q$ therefore enforces the correct asymptotic decay as a hard constraint. Additionally, the velocity components have different symmetry properties: $u'(x,y)$ is even in both $x$ and $y$, while $v'(x,y)$ is odd in both directions. Let $u_{pred}(x,y) = [u_1(x,y), u_2(x,y)]$ be the two network outputs. We enforce the required symmetry using:
\begin{equation}
    \begin{gathered}\label{eq:symmetric_constraint_uv}
    u' = u_1(x,y),\\
    v' = \frac{2xy}{x^2+y^2}\, u_2(x,y).
\end{gathered}
\end{equation}

\begin{figure}[t!]
	\centering
	\includegraphics[width=1\textwidth]{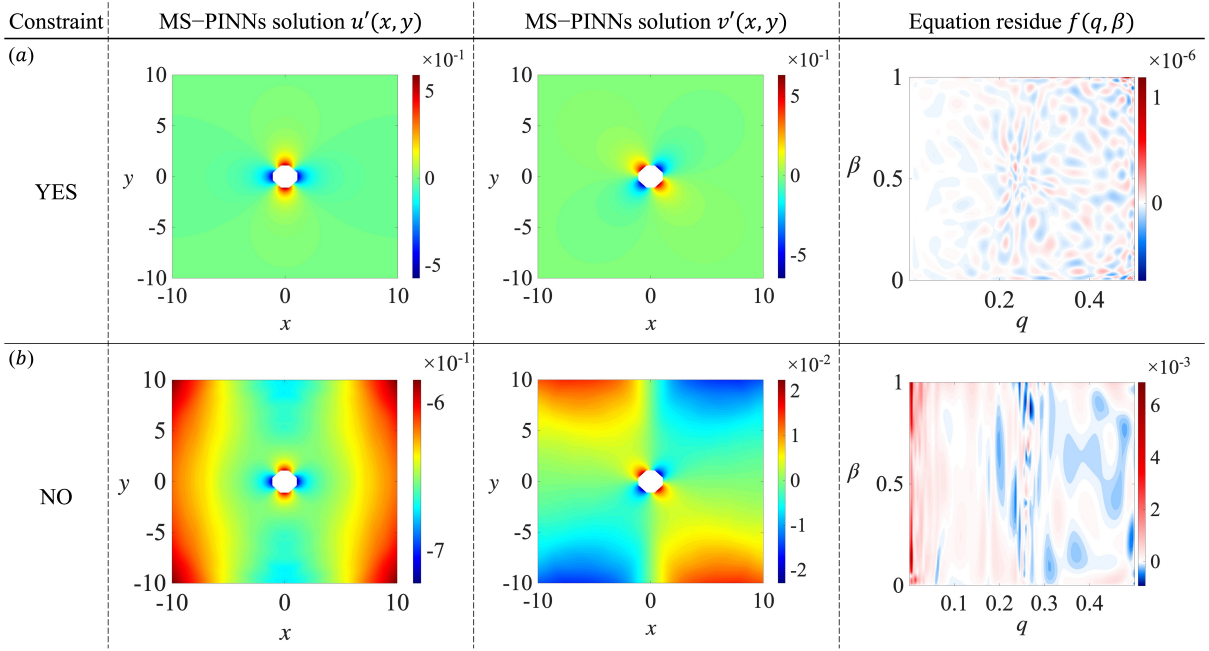}
	\caption{\label{fig:uv_result} {\bf MS-PINN solution of the velocity-based nonlinear equation.} ($a$) Predicted perturbation velocity fields $(u',v')$ and equation residuals of the velocity-based nonlinear equation~\eqref{eq:nonlinear_velocity_GE} for compressible flow at $M_\infty=0.2$, obtained under the hard constraint~\eqref{eq:uv_hard_constraint}. ($b$) Erroneous results obtained without applying the hard constraint.}
\end{figure}

Figure~\ref{fig:uv_result}($b$) shows a failure case when we solve equations~\eqref{eq:nonlinear_velocity_GE} and \eqref{eq:irrotationality_GE} without the hard constraint \eqref{eq:uv_hard_constraint} at $M_\infty=0.2$. Both $u'$ and $v'$ become nearly constant, and $v'$ is further confined around zero due to the symmetric constraint. The residual can decrease slightly, but it typically remains around $\mathcal{O}(10^{-3})$, with larger errors persisting in the infinite region. Figure~\ref{fig:uv_result}($a$) presents the final results for the nonlinear solution for compressible subsonic flow when the hard constraint is applied. However, the dual-equation formulation yields lower accuracy than the single-equation formulation. This limitation likely arises because the two governing equations do not provide sufficiently strong coupling to reinforce each other. Although dimensionality reduction is often considered a promising strategy in theory and especially in traditional numerical methods, in this problem, the higher-dimensional formulation unexpectedly leads to better performance within PINNs. Another possibility is that the standard PINN framework is already robust enough to handle the high-order derivatives in the potential formulation via automatic differentiation, so reducing the PDE order does not translate into a clear optimization advantage.

\section{Potential spectral artifacts}\label{sec:spectral artifact}

\begin{figure}[t!]
	\centering
	\includegraphics[width=1\textwidth]{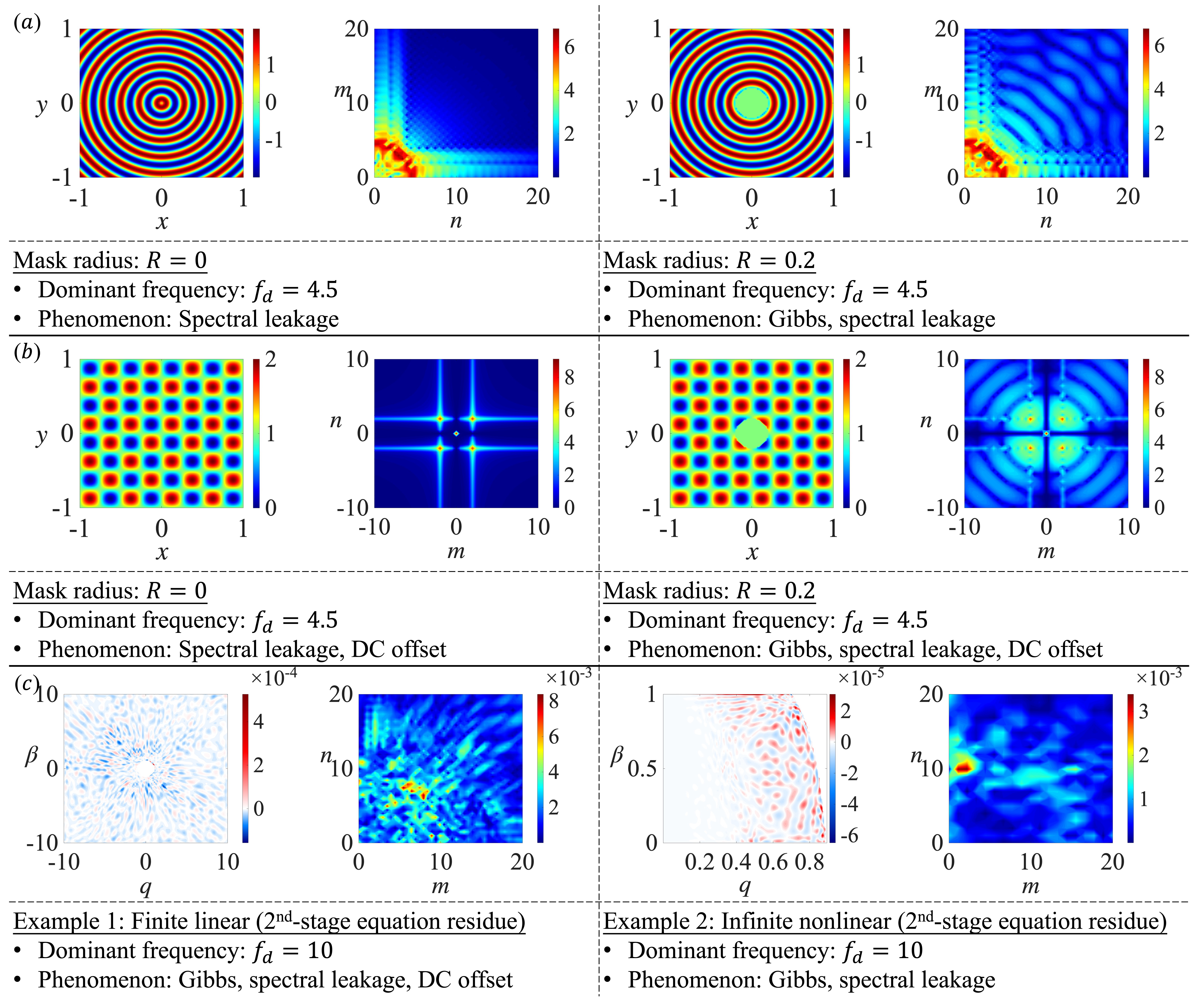}
	\caption{\label{fig:artifacts} {\bf Potential spectral artifacts.} ($a$) DFT results for $u=\sin(9\pi \sqrt{x^2+y^2})$ with and without constant-value masking in the inner region, together with the identified spectral artifacts. ($b$) Corresponding results for $u=\sin(4\pi x)\sin(4\pi y)+1$ under the same constant-value masking. ($c$) Equation errors from the second-stage training of the subsonic flow equation calculated over either a finite domain or an infinite domain. A summary of the dominant frequency, identified artifacts (e.g., DC offset, spectral leakage, Gibbs phenomenon) is provided in the accompanying table.}
\end{figure}

In Section~\ref{sec:MSPINN}, we introduce a general DFT-based approach for PDEs with irregular boundary conditions in MS-PINN training. This approach uses masking (setting values outside the physical domain to zero) so that the field can be analyzed on a regular grid. However, masking can introduce jump discontinuities at the mask boundary. These discontinuities may create oscillations near the boundary and produce artificial high-frequency components in the spectrum, which is the well-known Gibbs phenomenon. Figures \ref{fig:artifacts}($a$–$b$) demonstrate this effect with two examples, $u=\sin(9\pi \sqrt{x^2+y^2})$ and $u=\sin(4\pi x)\sin(4\pi y)+1$, under varying degrees of discontinuity. In both cases, a constant-valued masked region is imposed within a central circle of radii $0$ and $0.2$. Compared with the continuous case, the masked cases show a clear Gibbs phenomenon with additional high-frequency content. Similar artifacts can also appear when we mask the equation residuals. Figure~\ref{fig:artifacts}($c$) shows a representative spectrum from our study, in which pronounced high-frequency artifacts are present. These effects are common in spectral analysis and can usually be recognized and ignored without affecting the identification of the dominant frequency.

Moreover, windowing in a finite spatial domain can introduce low-frequency artifacts, a phenomenon known as spectral leakage. When the window function $w(x,y)$ - i.e., the constant-value masking in our case - is applied to the original signal $u(x,y)$, the Fourier transform becomes:
\begin{gather}\label{eq:leakage}
    U_w(f_x,f_y)=\mathcal{F}\{u(x,y)w(x,y)\}=U(f_x,f_y)*W(f_x,f_y),
\end{gather}
where the symbol $*$ denotes the 2D convolution. As a result, the spectral content is smeared: an originally single frequency spreads into adjacent frequencies, leading to an elevation of the low-frequency region in the spectrum. High-frequency components may also be artificially intensified. Figures~\ref{fig:artifacts}($a$–$b$) illustrate this effect, showing an increased magnitude in the low-frequency region with larger constant components. A similar phenomenon is evident in our study, as shown in Figure~\ref{fig:artifacts}($c$). Nevertheless, the dominant frequency remains clearly identifiable owing to its circular structure and relatively high magnitude.

Finally, a DC offset can appear when the (masked) field has a nonzero spatial mean, which produces an abnormally high spike at the origin of the Fourier spectrum. The Discrete Fourier Transform (DFT) is given by:
\begin{gather}\label{eq:DC}
    F[k,l] = \frac{1}{MN} \sum_{m=0}^{M-1} \sum_{n=0}^{N-1} f[m,n] \cdot e^{-j 2\pi \left( \frac{k}{M} m + \frac{l}{N} n \right)},
\end{gather}
and at $k=l=0$, we obtain $F[0,0]=\mathrm{mean}(f[m,n])\neq 0$. A nonzero mean therefore leads to a misleadingly large value at the origin. Figure~\ref{fig:artifacts}($b$) illustrates this effect by adding a constant shift of $1$, i.e., $u=\sin(4\pi x)\sin(4\pi y)+1$, which produces a pronounced spike at the origin of the spectrum. In practice, this spike can generally be disregarded as long as the dominant spectral components remain clearly identifiable.

\bibliographystyle{unsrt}
\bibliography{reference.bib}

\end{document}